\RequirePackage{fix-cm}
\documentclass{svjour3}                     
\smartqed  
\usepackage{graphicx}
\usepackage{comment,color}
\usepackage{txfonts}
\usepackage{color}
\usepackage{ulem}
\begin{document}

\title{ 
Insigths into the tribochemistry of silicon-doped carbon based films by 
{\it ab initio} analysis of water/surface interactions.
}

\author{Seiji Kajita         \and
        M. C. Righi
}


\institute{S. Kajita \at
              Toyota Central R\&D Labs., Inc., 41-1, Yokomichi,
              Nagakute, Aichi 480-1192, Japan \\
	      Istituto Nanoscienze, CNR-Consiglio Nazionale delle
	      Ricerche and Universita' di Modena e Reggio Emilia, I-41125 Modena,
              Italy\\
              \email{fine-controller@mosk.tytlabs.co.jp}           
              \and
              M. C. Righi \at
	      Istituto Nanoscienze, CNR-Consiglio Nazionale delle
	      Ricerche, I-41125 Modena, Italy \\
              \email{mcrighi@unimore.it}           
}

\date{Received: date / Accepted: date}

\maketitle

\begin{abstract}
Diamond and diamond-like carbon (DLC) are used as
coating materials for numerous applications, ranging from
 biomedicine to tribology.  Recently, it has been shown that 
 the hydrophilicity of the carbon films can be enhanced by silicon doping, 
 which highly improves their 
biocompatibility and
frictional performances. 
Despite the relevance of these properties 
 for applications, a microscopic understanding on the effects of silicon
is still lacking.
 Here we apply {\it ab initio}
 calculations to study the interaction of  water molecules with Si-incorporated
 C(001) surfaces.
 We find that the presence of Si dopants considerably increases the energy gain
 for water chemisorption and decreases the energy barrier for water dissociation
 by more than 50\%. We provide a physical rational for the phenomenon by 
 analysing the electronic
 charge displacements occuring upon adsorption.
 We also show that once hydroxylated, the surface is able to bind further water molecules
 much strongly than the clean surface via hydrogen-bond networks. This two-step process is
 consistent with and can explain the enhanced hydrophilic character 
observed in carbon-based films doped by silicon.

\end{abstract}

\section{Introduction}
Polycrystalline diamond and diamond like carbon (DLC) 
provide fully biocompatible, ultra-hard, low-friction, wear-resistant 
coatings, which make them very attractive for numerous applications 
ranging from biomedicine to tribology.
Recently, a number of studies has been devoted to the incorporation of
elements, such as silicon, nitrogen and metals
into the carbon films to modify their structural and surface
properties.\cite{sanchez,bewilogua,donnet,grill}
Silicon, in particular,  is incorporated in DLC to improve 
the adhesion strength on substrates,\cite{mori,miyake}
thermal stability,\cite{camargo,wu,varma,hatada}
biocompatibility\cite{zhao,bendavid} and reduce the friction
coefficient.\cite{oguri,kim,ikeyama,lanigan}

 The microscopic mechanisms that lead to the extremely low friction
coefficient of silicon-doped DLC (Si-DLC) are still unclear. 
It is expected that the presence of Si dopants modifies the
surface/water interaction and tribochemistry in humid environments.
In fact, silanol  groups (Si-OH) have been detected on the surface of
as-deposited Si-DLC in ambient air
by derivatization X-ray photoelectron spectroscopy.\cite{takahashi,kato} 
This surface termination, resulting from the dissociative
adsorption of water molecules, has been suggested 
to promote the formation of a water boundary layer,
which prevents the direct contact of the sliding surfaces
and consequently reduces friction.\cite{kato}
This hypothesis has been confirmed by
molecular dynamics (MD) simulations, which showed that 
water molecules strongly attached to the surface hydroxyl groups 
via hydrogen bond networks were able to separate the sliding
surfaces in shear conditions.\cite{washizu}

 A decrease of water contact angle with an
increase in Si doping in the DLC coating has been recently
reported,\cite{dearnaley,saito} 
 but the chemical role of the Si dopants has not been elucidated.
Understanding the microscopic mechanisms that govern the wettability of
doped DLC coatings is  of paramount importance not only for tribology,
but also for biomedical applications such as contact lens and the
artificial heart valves.\cite{borisenko,yi} 

Here we apply first principle calculations based on density functional
theory (DFT) to investigate the effects of silicon dopants on 
water adsorption at diamond surfaces.
Despite its relevance for lubrication,
biological materials application and microelectronics, the 
diamond/water interface has been scarcely investigated by 
{\it ab initio} methods.\cite{okamoto,manelli,jpcc-diamond,zilibotti}  
The use of a parameter-free approach is particularly important because
classical potentials hardly capture the water/surface interaction,\cite{galli}
 and
chemical reactions such as water dissociation at the surface cannot be
accurately described. 
We consider the diamond (001) surface, which 
presents a dimer reconstruction composed of carbon double bonds, to
resemble the thin DLC layer of sp$^2$ carbon often
detected in tribologial contacts.\cite{jpcc-diamond}
Silicon atoms are located at substitutional sites identified as most stable at the surface,
with a concentration typical for Si-DLC. 
The simulated crystalline system differs from the amorphous
matrix of sp$^2$ and sp$^3$ carbon of real Si-DLC coatings,\cite{grill,iseki,palshin}
however it constitutes a minimal model suitable to give insights into 
the {\it local} physical/chemical interactions 
that promote water
adsorption and dissociation at the Si sites. 
By comparing the adsorption energies and dissociation barriers of Si-doped
and clean diamond surfaces, we elucidate the chemical role of Si dopants
in modifying the hydrophilic character of the carbon-based films.

\section{Method}
We model the (001) surface with periodic supercells containing a diamond
slab reconstructed on both the sides with ten-layer thickness and (4$\times$3)
in-plane size. The surface presents a (2$\times$1)
reconstruction constituted of dimers that give rise to alternating rows
and trenches of sp$^2$- and sp$^3$-bonded carbon atoms. 
A vacuum region of 20 $\AA$ thickness is included in the supercell to separate each slab from its
periodic replicas along the [001] direction. 
Silicon atoms are located at substitutional sites, 
consistently with the experimental observation that silicon atoms
in Si-DLC are surrounded by carbon atoms, and not by oxygen or other
silicon atoms.\cite{iseki,palshin} 
A 8.3\% Si concentration is realized, which is consistent with
that used in the experiments.\cite{nakanishi}

Static {\it ab initio} DFT calculations are performed by means of
the pseudopotential/plane-waves computer code included in the QUANTUM
ESPRESSO package.\cite{qe}
The Perdew, Burke and Ernzerhof generalized gradient
approximation (GGA-PBE) is used for the exchange-correlation functional,\cite{pbe} 
and all the simulations include the spin polarization. 
The electronic wave
functions are expanded on a plane-waves basis set with a cutoff energy
of 25 Ry, and the ionic species are described by ultra-soft pseudo
potentials.\cite{vanderbilt} 
The $k$ points in the Brilliouin zone are sampled by means of a $2 \times 3 \times 1$
Monkhorst-Pack grid.\cite{monkhorst}
In the geometry optimization procedure,
all the atoms except for those belonging to the slab bottom layer are
allowed to relax until the forces become lower than 0.001 Ry/Bohr. 
The adsorption energy of a water molecule on the surface ia calculated as
$E_{ad} = E_{tot} - E_{surf} - E_{H_2O}$, where $E_{tot}$, $E_{surf}$  
and $E_{H_2O}$ are the energies
of the adsorbate system,
the clean surface and isolated water molecule. The reaction path and energy barrier for water
dissociation are calculated by means of the nudged elastic band (NEB) 
method within the "climbing the image" scheme.\cite{cineb1,cineb2}
The NEB method is able to identify the minimum
energy path (MEP) for the system transition between two stable
configurations. 
In our case they correspond to water physisorption and
dissociative chemisorption configurations, respectively. 
In the NEB calculations, 
we reduce the slab thickness to six layers in order
to accelerate the convergence of the method.
\section{Results and Discussions}
\subsection*{Effects of silicon dopants on the process of surface hydroxylation.}

\begin{figure}[htbp]
 \begin{center}
 \includegraphics[width=0.95\linewidth]{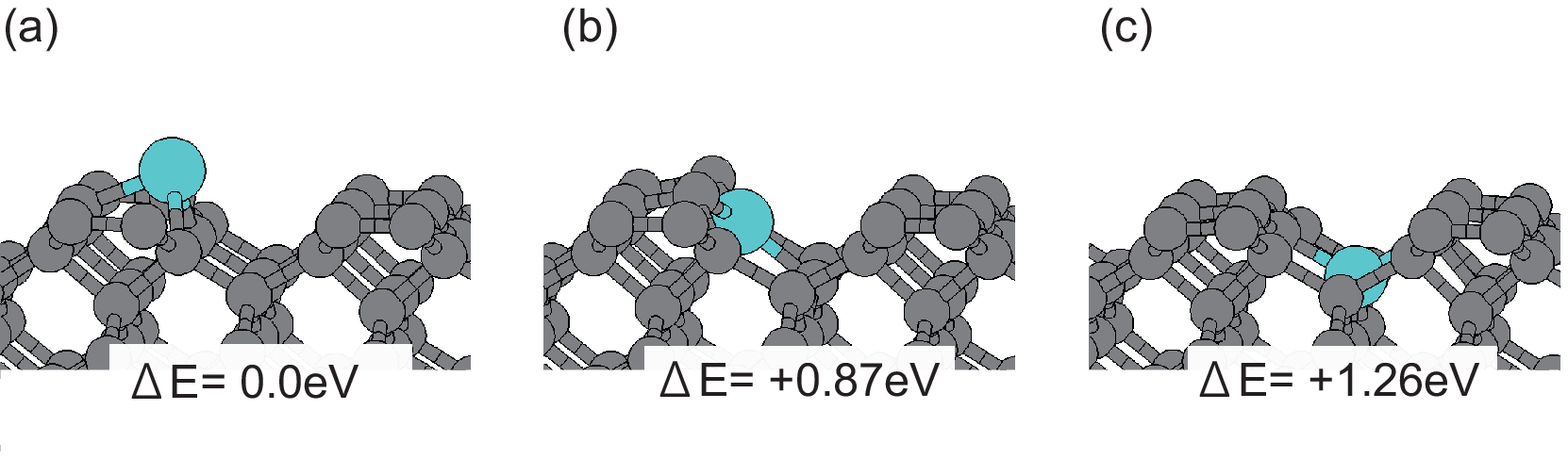}
  \caption{ Relative stability of different substitutional sites
  for the Si dopant. 
  The energy differences are referred to the most stable configuration,
  (a), where the Si atom is loacated in the first surface layer and
  forms a Si-C heterodimer.
The grey and blue balls indicate Si and C atoms, respectively.
}  　\label{pic:si-configurations}
 \end{center}
\end{figure}

In order to construct a realistic model for the Si-incorporated
C(001) surface, we identify the most favorable location for the silicon atoms
by comparing the stability of different substitutional sites. As can be
seen in Fig. \ref{pic:si-configurations},
the compared sites are located within the first, second and third surface
layers.  The configuration containing the Si-C heterodimer 
(Fig. \ref{pic:si-configurations}(a))
turned out to be the most energetically favored. 
Indeed, Si dopant surrounded by C-sp$^2$ atoms
has been identified as the prevailing configuration in Si-DLC films
by nuclear magnetic resonance experiments. \cite{iseki}

\begin{figure}[htbp]
 \begin{center}
 \includegraphics[width=0.9\linewidth]{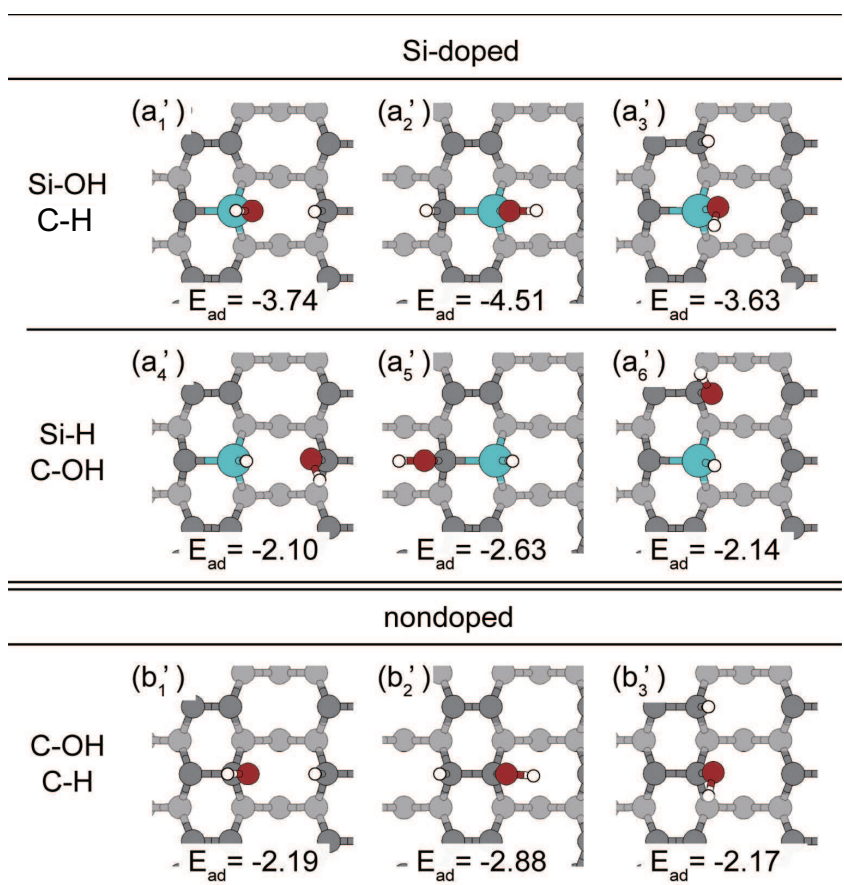}
  \caption{ 
Dissociative adsorption configurations for a water molecule on the 
Si-incorpored and clean C(001) surfaces.
The adsorptions energies, $E_{ad}$, are reported in eV.
The red and white balls indicate O and H atoms, respectively.
}\label{pic:table}%
 \end{center}
\end{figure}

As a next step, we evaluate the effects of silicon on the dissociative
adsorption of water molecules. The considered adsorption configurations are modeled by
positioning the two fragments (-H and - OH) of the dissociated molecule
on different sites. In particular, we model adsorption processes
involving either only one dimer or two adjacent dimers (belonging both
to the same row or separated by a trench). The optimized configurations
and the corresponding adsorption energies are shown in Fig.
\ref{pic:table}. 
As can be seen from the first and second rows, where the Si-doped surface
is considered, the adsorption configurations
containing a Si-OH group are more stable than those containing a Si-H
group by more than $1$ eV. The comparison between the adsorption
energies for Si-doped and undoped surfaces (first and third rows of
Fig. \ref{pic:table}) indicates that the stability of a Si-OH group is significantly
higher than that of a C-OH group by approximately $1.5$ eV. 
On the contrary, the energy gain for hydrogen adsorption is almost independent from
the presence of silicon (second and third rows). These results indicate
that the Si dopant selectively stabilizes the hydroxyl group.

To better understand the above described results, we analyze the
electronic charge displacements and the projected density of states
(PDOS) of hydroxylated surfaces.

\begin{figure}
 \begin{center}
 \includegraphics[width=0.8\linewidth]{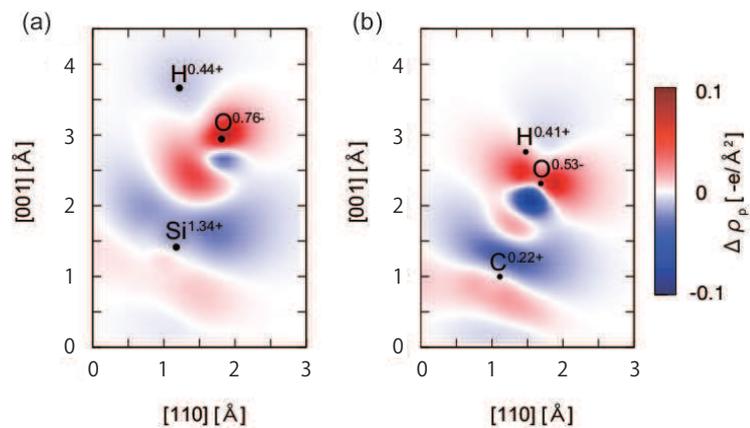}
  \caption{
Charge displacements projected on the ($\bar{\textrm{1}}$10) plane upon adsorption of OH
  fragments on the (a)  Si-doped and (b) undoped surfaces. 
The  superscript of each element indicates the partial charge obtained by
  the L$\ddot{\textrm{o}}$wdin charge populations analysis.
}\label{pic:dcharge-oh}
 \end{center}
\end{figure}

The charge displacement, $\Delta \rho$, occurring upon adsorption of an OH
fragment on the surface calculated as 
$\Delta \rho = \rho_{tot} - \rho_{surf} - \rho_{frag}$, 
where 
$\rho_{tot}$,  $\rho_{surf}$ and  $\rho_{frag}$
are the electronic density of the
adsorbed system, the clean surface and isolated fragment. 
Figure \ref{pic:dcharge-oh} shows the
projected charge displacements occurring upon hydroxyl adsorption on the
Si-doped (a) and undoped (b) surfaces. 
The projection of $\Delta \rho$  on the ($\bar{\textrm{1}}$10) plane is obtained
by integration along the perpendicular direction. 
An accumulation of electrons can be observed
both in the O-Si and O-C bond regions, consistently with the
covalent character of these bonds. 
The oxygen bound to the Si
site attracts more electrons than the oxygen bound to the C site,
as  confirmed by the analysis of the
L$\ddot{\textrm{o}}$wdin charges 
displayed in Fig. \ref{pic:dcharge-oh}. \cite{lowdin1}
The O-Si bond presents, thus, an ionic character more pronounced than the O-C
bond.

\begin{figure}
 \begin{center}
 \includegraphics[width=0.65\linewidth]{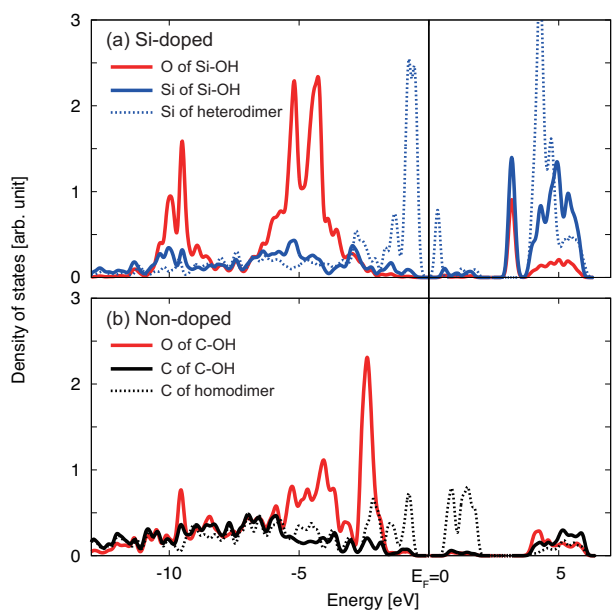}
  \caption{ PDOS of the OH-adsorbed 
on the (a)  Si-doped and (b) undoped surfaces. 
The origina of the horizontal axis corresponds to the Fermi energy.
}  　\label{pic:pdos-oh}
 \end{center}
\end{figure}

To establish a connection between the bonding character and the
adsorption energy, we analyze the PDOS for hydroxyl adsorptions shown 
in Fig. \ref{pic:pdos-oh}.
On the clean surfaces, the silicon atom in the heterodimer and
the carbon atom in the homodimer (dashed lines) have distinctive peaks
below their Fermi energies, at about $-0.5$ eV. The magnitude of the Si
surface state is higher than that of the undoped surface. 
When a hydroxyl adsorbs on the surface, these surface states disappear
because new bonds are formed with the oxygen atom. The occupied surface state
of the Si dopant before the adsorption (dashed blue line) shifts above
the Fermi energy as empty states (solid blue line), which is an evidence
that electrons of the Si dopant transfer to the adsorbed oxygen atom and
a bond with enhanced ionic character is formed. Although the electron
transfer occurs also in the case of the undoped surface, 
the oxygen peak of the Si-OH group is located in a lower
energy region than that of the C-OH group (red lines). 
This suggests that the transfer of electrons from the dopant to the oxygen 
stabilizes the hydroxyl adsorbate. 
Such enhanced electronic
transfer is due to the lower electronegativity of silicon: 
O (3.44) $>$ C (2.55) $>$ Si (1.90), 
where the numbers in the parentheses are the Pauling
values.\cite{en}
The lower electronegativity of the Si atoms, which 
are surrounded by carbon atoms in Si-DLC,  thus controls the
bond polarization and promotes the surface hydroxylation,
as shown by the reaction paths described in the following.

\begin{figure}[htbp]
 \begin{center}
 \includegraphics[width=0.75\linewidth]{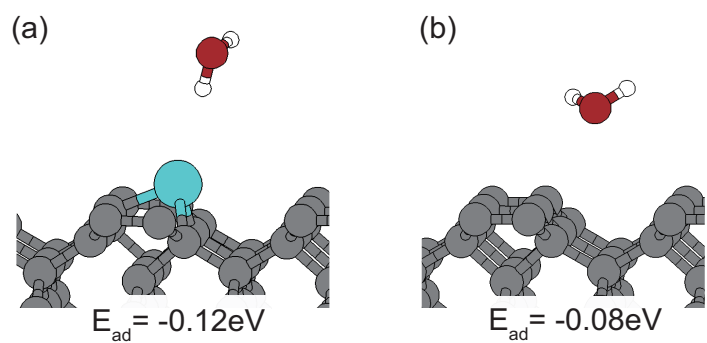}
  \caption{ Most stable configurations identified for molecular adsorption on
(a) Si-doped and (b) undoped surfaces.
}  　\label{pic:physisorption}
 \end{center}
\end{figure}

\begin{figure}
 \begin{center}
 \includegraphics[width=1.0\linewidth]{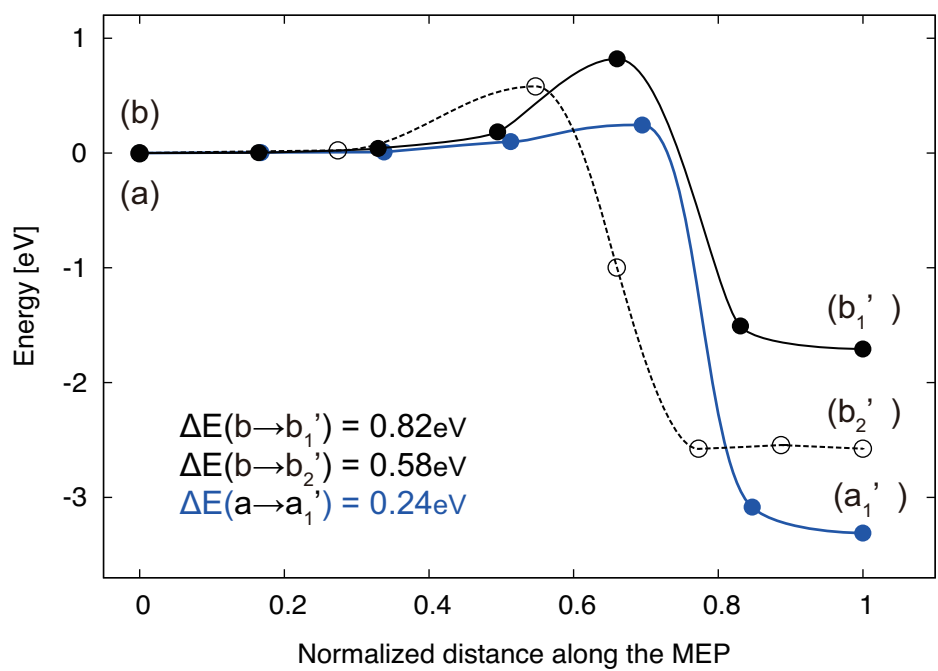}
  \caption{ Reaction paths and the energy barriers $\Delta E$
 for water dissociations obtained by NEB calculations. 
The blue and black plots indicate the NEB images for the Si-doped and
  undoped surfaces, respectively.
The initial (final) states of the reaction paths correspond to the configurations shown in 
Fig. \ref{pic:physisorption} (Fig. \ref{pic:table}), which are labeled
  in the same way.
} \label{pic:neb}
 \end{center}
\end{figure}

We identify the reaction path and energy barrier for water
dissociation both on the clean and Si-incorporated C(001) surfaces, 
by applying the NEB method. 
As initial states for the reactions we consider
the most favorable configurations for water adsorption
shown in Figs. \ref{pic:physisorption} (a) and (b). \cite{phys_detail}
The corresponding final states are chosen 
 among the configurations of Fig. \ref{pic:table}.
In particular, we consider dissociation both on the trench 
(a$'_1$, b$'_1$) and on the dimer (a$'_2$, b$'_2$).
In the case of Si-doped surface, the minimum energy path (MEP) 
identified by the NEB algorithm for the a $\rightarrow$ a$_2'$ transition
displays an intermediate minimum corresponding to the a$'_1$ state.
This means that one water molecule physisorbed with
the geometry of Figs. \ref{pic:physisorption} (a) hardly reacts with only one dimer 
to form the a$'_2$ configuration.
The latter is most likely obtained 
by the dissociation of more than one molecule.
Figure \ref{pic:neb}  summarized the results 
obtained for the reaction paths and energy barriers for water
dissociation on the Si-doped surface (blue line) and the undoped surface
(black lines). In the latter case the comparison between the results
obtained for the 
b $\rightarrow$ b$_1'$ and
b $\rightarrow$ b$_2'$ 
transitions indicate that water
dissociation occurs most likely within the dimer row rather than 
within the trench, in agreement with previous calculations.\cite{manelli}
The comparison between the a $\rightarrow$ a$_1'$ 
and b $\rightarrow$ b$_2'$ paths reveals that the
Si dopant reduces the energy barrier for the water dissociation by more
than 50\%.
This significant reduction of the reaction barrier is caused by the 
stabilization of the hydroxylated configuration by the Si dopant.
The reaction energy associated to the a $\rightarrow$ a$_1'$ transition
is, in fact, 
0.75 eV higher than the reaction energy of the 
b $\rightarrow$ b$_2'$ transition.

\subsection*{Effects of surface hydroxylation on water adsorption}

\begin{figure}[htbp]
 \begin{center}
 \includegraphics[width=0.95\linewidth]{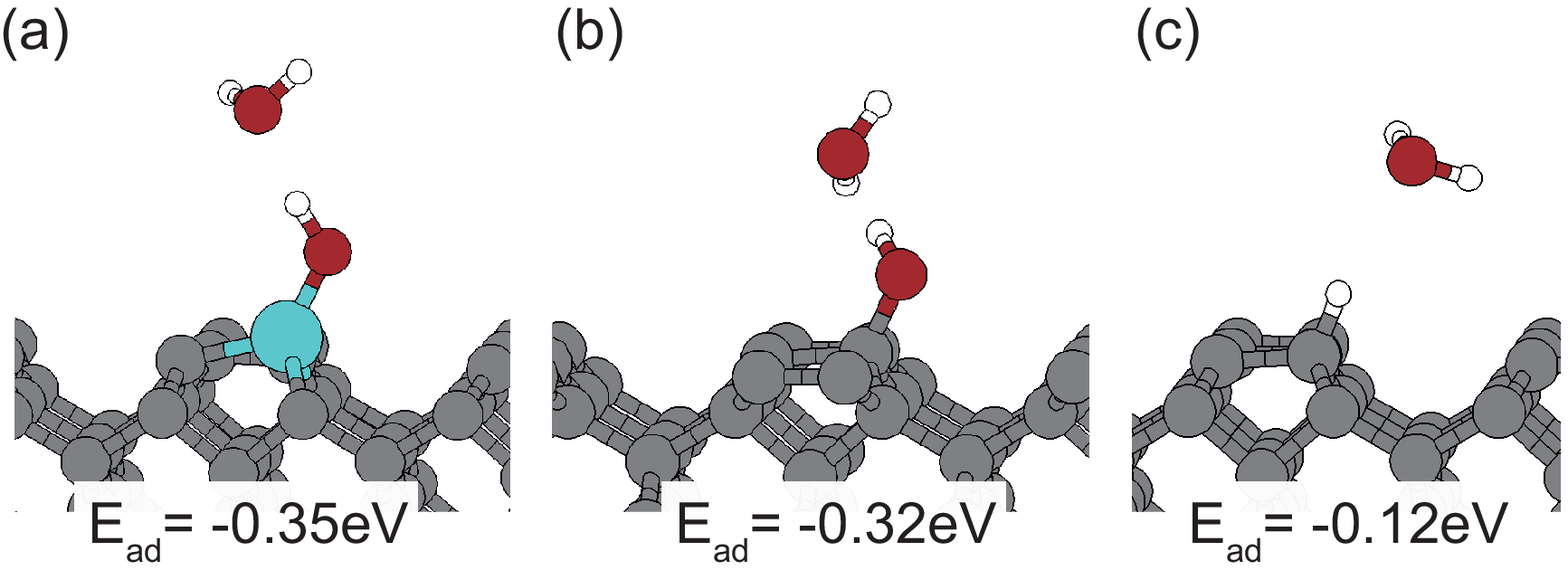}
  \caption{ Water physisorption at OH-and H-terminated
  surface sites. In the first case both
  Si-doped and undoped surfaces are considered.
}  　\label{pic:water_adsorptions}
 \end{center}
\end{figure}

As last step in our analysis, we evaluate the effects of surface
hydroxylation on the physisorption of further water molecules,
i.e. on the surface wettability.
To this aim we compare the energy for molecular
adsorption on OH-terminated, 
H-terminated and clean surfaces. In the first case we consider
both Si-doped and undoped surfaces.
The calculated adsorption energies, which are reported
in Figs. \ref{pic:physisorption} and \ref{pic:water_adsorptions},
indicate that the surface hydroxylation highly enhances water
attraction: the physisorption energy 
in the presence of adsorbed hydroxyl groups
is three times higher than in the absence. 
Furthermore, the distance between the physisorbed molecule 
on the hysroxylated surface (calculated
as the distance between the O atom of the physisorbed molecule 
and the H atom of the adsorbed -OH group)
ia of about 1.8~{\AA}.
This value is lower than that obtained for the hydrogenated surface of Fig. \ref{pic:water_adsorptions}(c)  and
is very close to 1.9~{\AA}, which is the typical length of hydrogen-bonds in bulk water.
These results indicate that the surface 
hydroxylation enhances 
the attraction of water molecules towards 
the surface thanks to the
formation of hydrogen bond networks that involve the chemisorbed -OH
groups. 
This increases the diamond/DLC hydrophilicity.

It should be kept in mind that the GGA functional in DFT fails in the exact
prediction of adsorption energies
because of the inadequate description of the van der Waals (VdW) interactions.
\cite{manelli,kajita_tio2,sprik,johnson}
However, 
the qualitative conclusions derived from this study should hold
on the basis of a
comparative analysis of the result present in the literature:
In our previous DFT study the energies
for water physisorption and dissociation on the diamond
surface were calculated both with and without the inclusion
of VdW interactions. The results
indicate that VdW inclusion increases the water
physisorption energy 
approximately by $0.1$ eV on average and decreases the
energy barrier for water dissociation by 30\%. \cite{manelli} 
Even if we take into account of the errors maximally,
the conclusions on the effects of Si dopants here 
derived will not be altered because they are based on
estimated energy differences that are much higher than these errors.

Finally, by comparing the adsorption energies
reported in Figs. \ref{pic:water_adsorptions}(a) and (b),
we observe that water adsorption 
on hydroxylated surfaces is not significantly affected by the presence
of Si dopants. Therefore, the main effect of Si dopants is to
favor water dissociation.
The consequent hydroxyl-termination
enhances the hydrophilic character of the surface.
 
\section{Conclusions}

To shed light into the effects of Si dopants in increasing 
the hydrophilicity of carbon-based coatings, we study the interaction
of water molecules with Si-doped and undoped C(001) surfaces
by means of {\it ab initio} calculations.
We find that the energy gain for water dissociative adsorption 
increases by approximately 1.5 times in the presence of 
substitutional Si atoms at the surface. 
The analysis of the electronic charge displacements occurring upon
adsorption reveals that this stabilization effect 
is due to the larger polarization of Si-OH bonds with respect
to C-OH bonds.
We apply the NEB method to identify the reaction paths 
for water dissociation and find that 
the energy barrier on the Si-incorporated surface
is 50\% lower than on the clean surface.
The process of surface
hydroxylation is, thus, highly favored
by Si-dopants.
 
Once hydroxylated, the surface is able to attract further water
molecules more strongly than the clean surface by
hydrogen bond formation (the 
 energy is more than three times higher on the
OH-terminated surface than on the clean surface). 
These two-step tribochemical process is consistent with and can explain the 
enhancements of hydrophilic character  and the consequent friction reduction
observed in carbon films doped by silicon.
\begin{acknowledgements}
We acknowledge the CINECA consortium for the availability of high
performance computing resources and support through the ISCRA-B TRIBOGMD
project.
\end{acknowledgements}


\begin{thebibliography}{9}
\bibitem{phys_detail}
The latter are identified by comparing the
water adsorption energy at different locations. 
In particular, we sample six positions on a homogeneous grid along the
[110] direction centered on the Si atom and covering the heterodimer
and half of the sp$^3$ trench. The in-plane coordinates of the oxygen atom are fixed
at the grid point, and all the other degrees of freedom are fully
relaxed except for the slab bottom layer.
\end{thebibliography}
\end{document}